\documentclass[epj,pdftex]{webofc}
\usepackage[varg]{txfonts}   
\woctitle{XLVI International Symposium on Multiparticle Dynamics}
\begin{document}
\title{Search for Muonic Dark Forces at BABAR}
%
%

\author{Romulus Godang\inst{1}\fnsep\thanks{\email{godang@southalabama.edu}}\\
On Behalf of the BABAR Collaboration}

\institute{
Department of Physics\\ 
University of South Alabama\\
411 University Boulevard, North\\
Mobile, AL 36688}

\abstract{%
Many models of physics beyond Standard Model predict the existence of light Higgs states, dark photons, 
and new gauge bosons mediating interactions between dark sectors and the Standard Model. 
Using a full data sample collected with the BABAR detector at the PEP-II $e^+e^-$ collider, we report 
searches for a light non-Standard Model Higgs boson, dark photon, and a new muonic dark force mediated 
by a gauge boson ($Z'$) coupling only to the second and third lepton families. Our results significantly 
improve upon the current bounds and further constrain the remaining region of the allowed parameter space.}
\maketitle
\section{Introduction}
Many astrophysical and cosmological observations indicate that a fraction of energy density in the universe is due 
to non-baryonic matter. The microscopic nature of dark matter is currently unknown. Models of physics beyond Standard Model 
predict the existence of a new non-Abelian gauge group Higgs with gauge boson masses below 10 GeV~\cite{birkedal}.  
The WIMP hypothesis suggested that dark matter is assumed to consist of stable particle with low masses. Such new gauge 
bosons can typically interact with other Standard Model elementary particles. The new gauge boson, $Z'$, can couple to 
the Standard Model field. SM fields can be directly charged under new gauge boson or the $Z'$ may couple with the SM hypercharge 
boson~\cite{holdom}.  

Based on the $L_\mu - L_\tau$ model~\cite{altmannshofer} one of the most promising candidates based on gauging the existing approximate global 
symmetries of the Standard Model (SM) is the gauge group associated with the difference between muon and tau-lepton number. 
The gauge $L_\mu - L_\tau$ model portal to the $Z'$ has all features of being coupled only to the leptons of the second and third generation.  
The $L_\mu - L_\tau$ model has been studied in the neutrino mass model. It has also been studied in the (g-2) current discrepancy. Consequently it was found 
that the tentative explanation of the (g-2) anomaly in models with large electron coupling is now excluded at least for multi-GeV and heavier $Z'$ bosons. 
The model explains the possible production of $Z'$ gauge boson production via $e^+ e^- \to \to \mu^+ \mu^- Z'$, $Z' \to \mu^+ \mu^-$ that primarily comes 
from the radiation of heavy-flavor leptons and is shown in the following Fig.~\ref{fig-0}.
\begin{figure}
\centering
\includegraphics[width=5cm,clip]{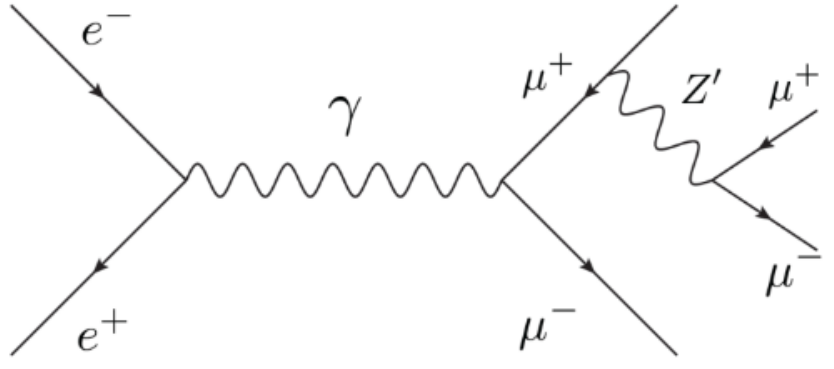}
\caption{Feynman Diagram for production of gauge boson $Z'$ based on the $L_\mu - L_\tau$ model at an $e^+ e^-$ collider.}
\label{fig-0}       
\end{figure}

\section{Data and Event Selection}
We used the data collected by the BABAR detector~\cite{babar_det} with the total luminosity of 514 fb$^{-1}$. Most of the data 
were taken at the $\Upsilon(4S)$ resonance plus including about 28 fb$^{-1}$ data at $\Upsilon(3S)$ and 14 fb$^{-1}$ data 
at $\Upsilon(2S)$ and 48 fb$^{-1}$ data at the off-resonance. The $\Upsilon(4S)$ resonance decays to a pair of 
$\bar{B}B$~\cite{babar05}. We used about 5\% of the data set to validate and optimize the analysis method. The rest of the data 
was only examined after finish finalizing the analysis method. For the background study we generated signal Monte Carlo (MC) samples.  

Signal MC events are generated using MadGraph 5~\cite{madgraph5}, which calculates matrix elements for the sample. The MC then 
were showered using Pythia 6~\cite{pythia6} for about 30 different $Z'$ mass hypotheses. The main background comes from the
QED processes. We generate the direct processes of $e^+ e^- \to \mu^+ \mu^- \mu^+ \mu^-$ using Diag36~\cite{diag36}, which includes
the full set of the lowest order diagrams. The Diag36 does not include initial state radiation (ISR) samples. The events 
of the process of $e^+ e^- \to e^+ e^- (\gamma)$ are generated using BHWIDE~\cite{bhwide} and the MC events of 
$e^+ e^- \to \mu^+ \mu^- (\gamma)$ and $e^+ e^- \to \tau^+ \tau^- (\gamma)$ are generated using KK~\cite{kk}. The off-resonance 
data samples, $e^+e^- \to \bar{q}q$ (q = u, d, s, c), are simulated using EvtGen~\cite{evtgen}. The events processes of
$e^+e^- \to \psi(2S)\gamma$ then $\psi(2S) \to \pi^+ \pi^- J/\psi$ and $J/\psi \to \mu^+ \mu^-$ were generated using a structure 
function technique~\cite{arbuzov,caffo}.  Finally the detector acceptance and reconstruction efficiency are determined 
using MC simulation based on GEANT4~\cite{geant4}.

\section{$Z'$ Measurement}
We select events containing exactly two pairs of oppositely charged tracks, consistent with the topology of the process: 
$e^+ e^- \to \mu^+ \mu^- Z'$ and $Z' \to \mu^+ \mu^-$ final state. The muons are identified by particle identification
algorithms for each track. We require the sum of energies of the electromagnetic clusters that are not associated 
to any track must be less than 200 MeV. We finally reject events that come from the $\Upsilon(3S)$ and $\Upsilon (2S)$,
where $\Upsilon(2S, 3S) \to \pi^+\pi^-\Upsilon(1S)$, $\Upsilon(1S) \to \mu^+ \mu^-$ decays if the dimuon combination is within
100 MeV of the $\Upsilon(1S)$ where pions are misidentified as muons. 

\begin{center}
\begin{figure}
\begin{tabular}{lr}
\includegraphics[width=6.6cm]{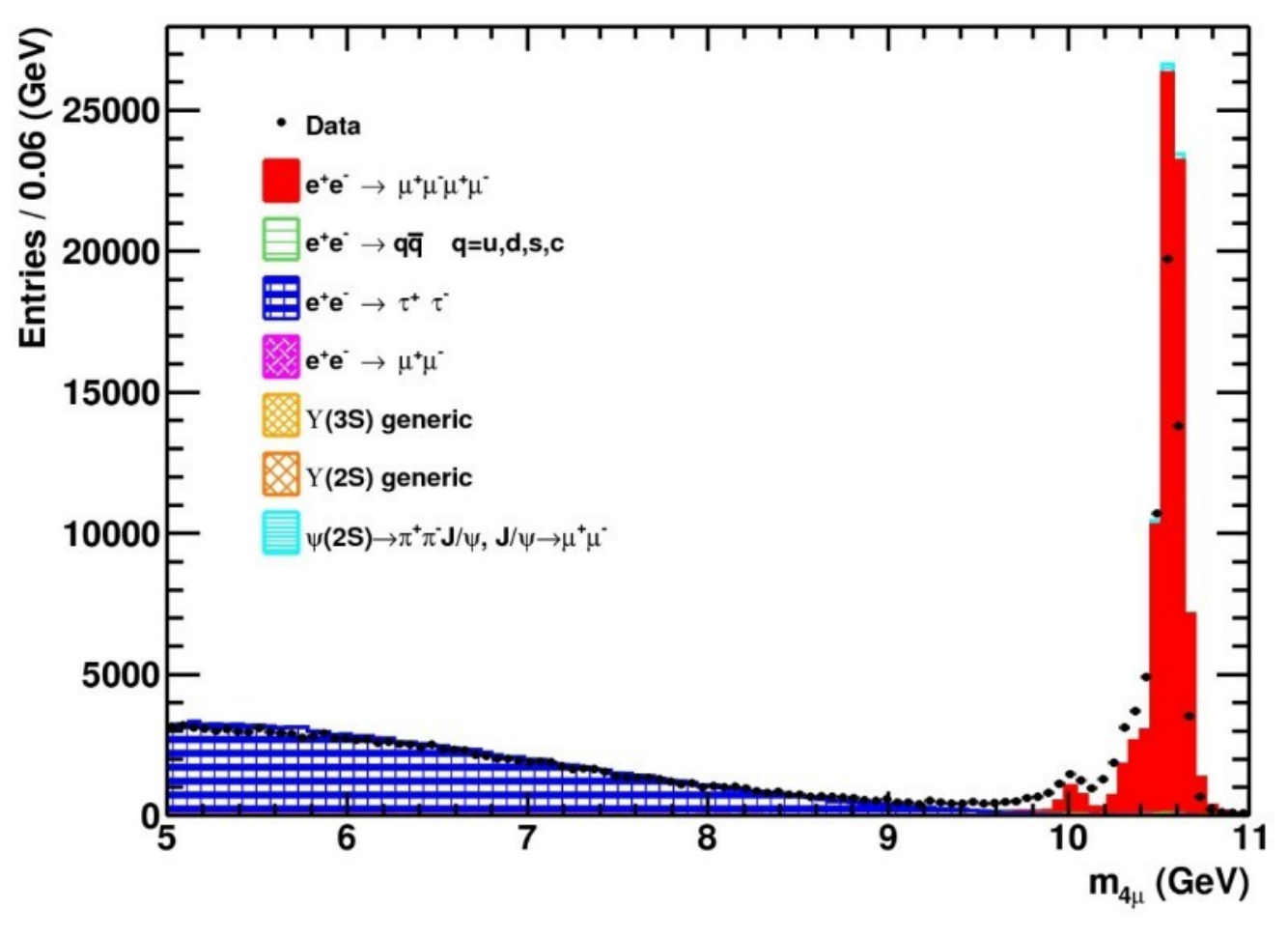}\hspace{0.4cm} 
\includegraphics[width=6.6cm]{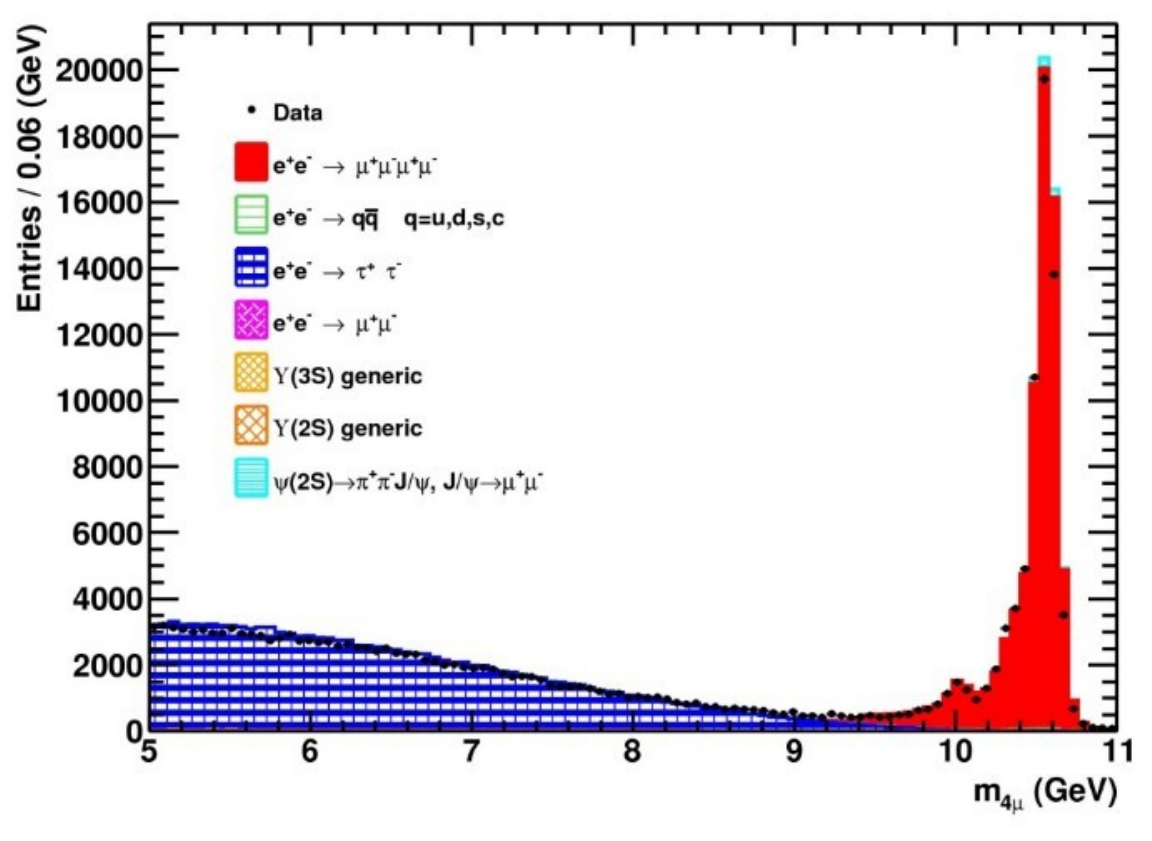}
\end{tabular}
\caption{(Left) The four-muon invariant mass distribution with the Monte Carlo predictions of various processes
including direct decay of $e^+ e^- \to \mu^+ \mu^-\mu^+ \mu^-$ normalized to the data luminosity. 
(Right) The four-muon invariant mass distribution with the Monte Carlo predictions of various processes by allowing
the initial state radiation emission.}
\label{fig-1}
\end{figure}
\end{center}

The distribution of the four-muon invariant mass after all selections is shown in Fig.~\ref{fig-1} (left). At the low mass of the 
four-muon invariant mass, $m(4\mu) < 9$ GeV, is well reproduced by the Monte Carlo simulation including direct decays of $e^+ e^- 
\to \mu^+ \mu^-\mu^+ \mu^-$, however, the Monte Carlo simulation overestimates the full energy peak by $\sim 30\%$ and fails to reproduce
the radiactive tail. The overestimate simulation is expected because the Diag36 simulation does not simulate the initial state radiation (ISR)
events. To estimate the potential ISR emission we select $e^+ e^- \to \mu^+ \mu^-\mu^+ \mu^-$ events by requiring a four-muon invariant mass
distribution within 500 MeV of the nominal center-of-mass energy. We also require the tracks to originate from the interaction point to within its
uncertainty and constraining the center-of-mass energy of the system to be within the beam energy spread. The four-muon invariant mass after allowing
the potential ISR emission is now fit as shown in Fig.~\ref{fig-1} (right).      

\begin{figure}
\centering
\includegraphics[width=8.6cm,clip]{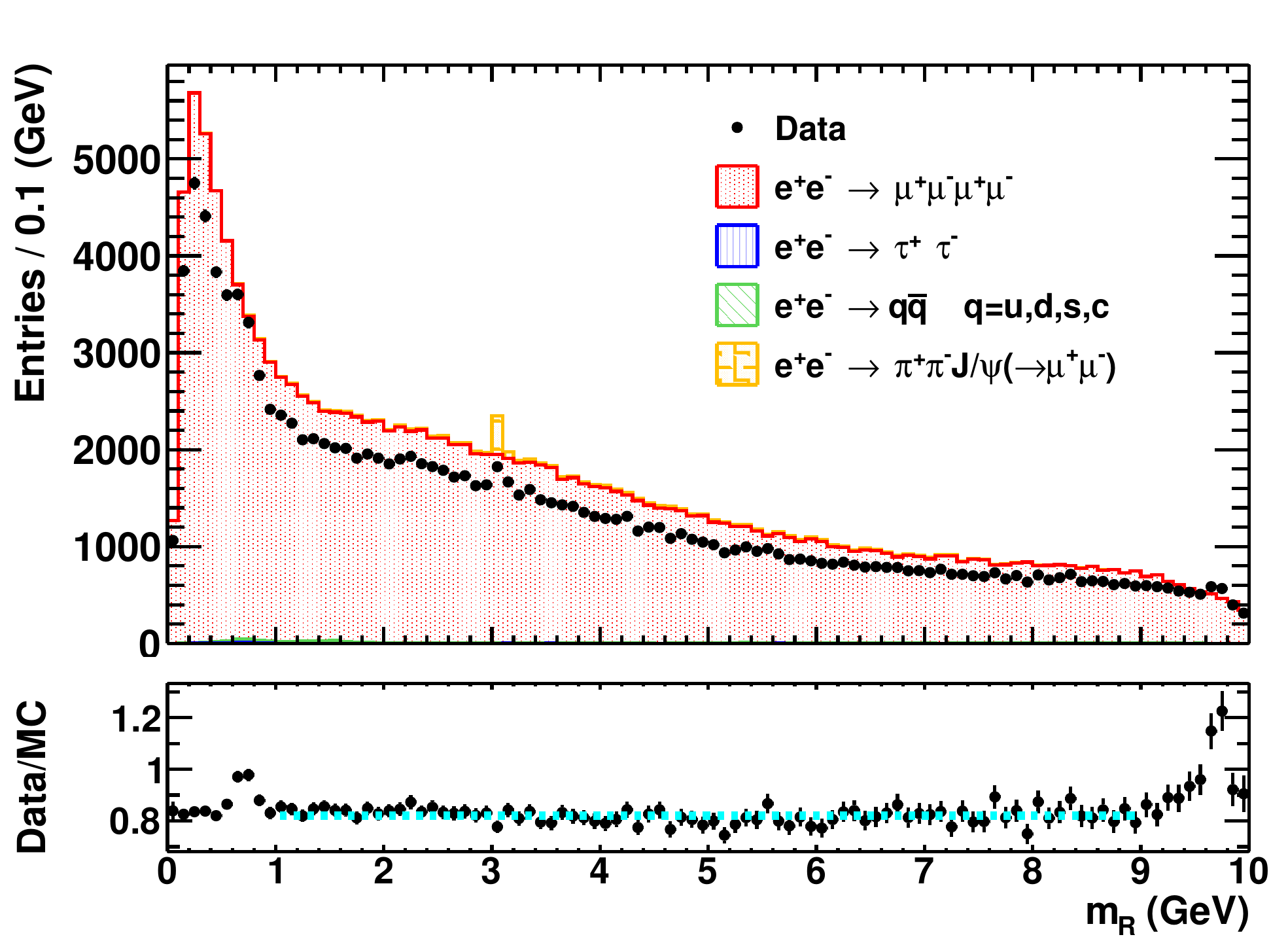}
\caption{The distribution of the reduced dimuon mass, $m_R = \sqrt{m^2_{\mu^+ \mu^-} - 4m^2_\mu}$, for the optimization
sample after applying all selections, together with the Monte Carlo predictions of various processes with normalized to the data 
luminosity. The fit ratio between the reconstructed and simulated events is shown as a blue dashed line.}
\label{fig-2a}       
\end{figure}

\begin{figure}
\centering
\includegraphics[width=8.6cm,clip]{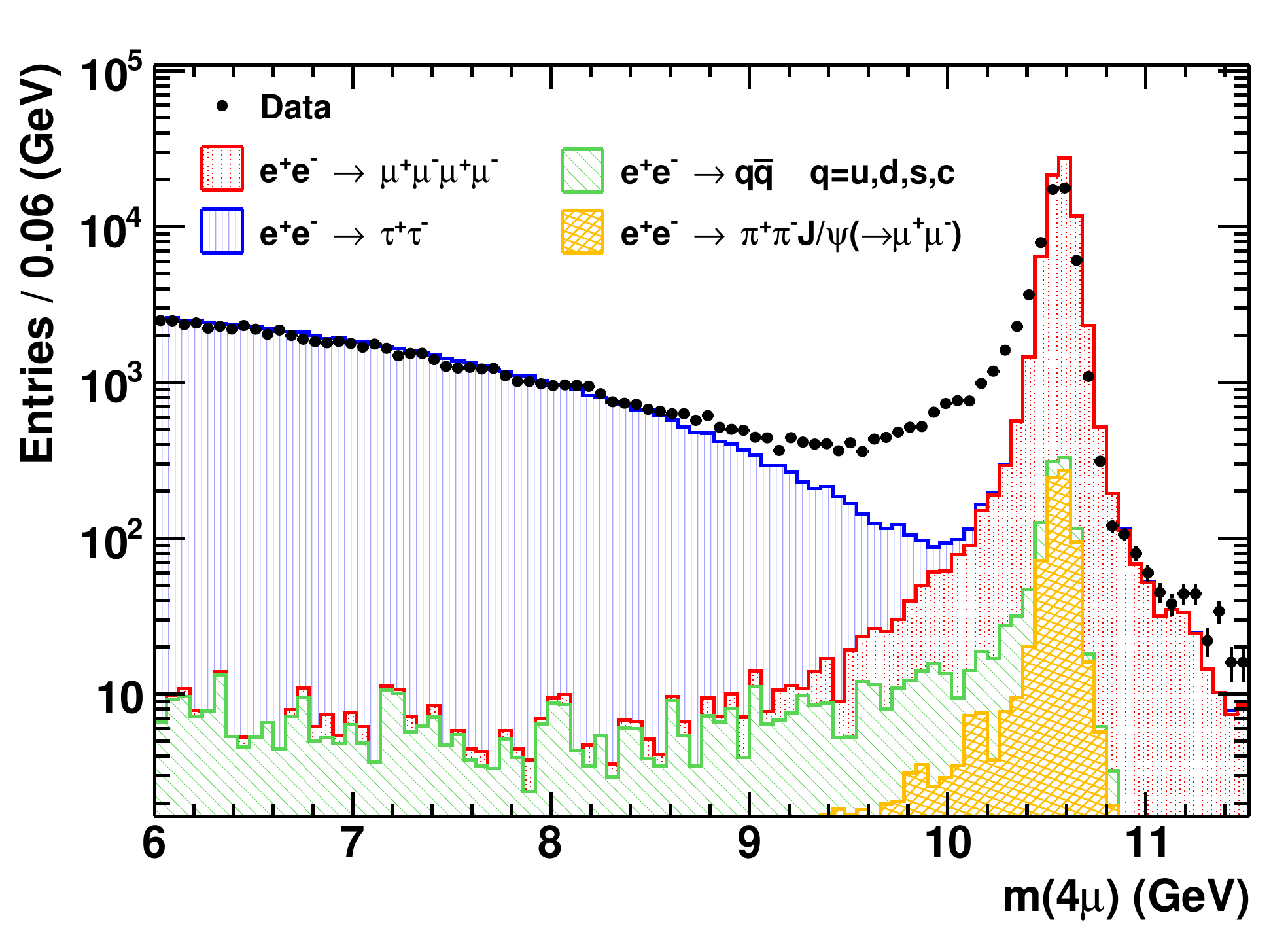}
\caption{The distribution of the reduced dimuon mass, $m_R = \sqrt{m^2_{\mu^+ \mu^-} - 4m^2_\mu}$ in log scale together
with the Monte Carlo predictions of various processes.}
\label{fig-2b}     
\end{figure}

\begin{figure}
\centering
\includegraphics[width=8.6cm,clip]{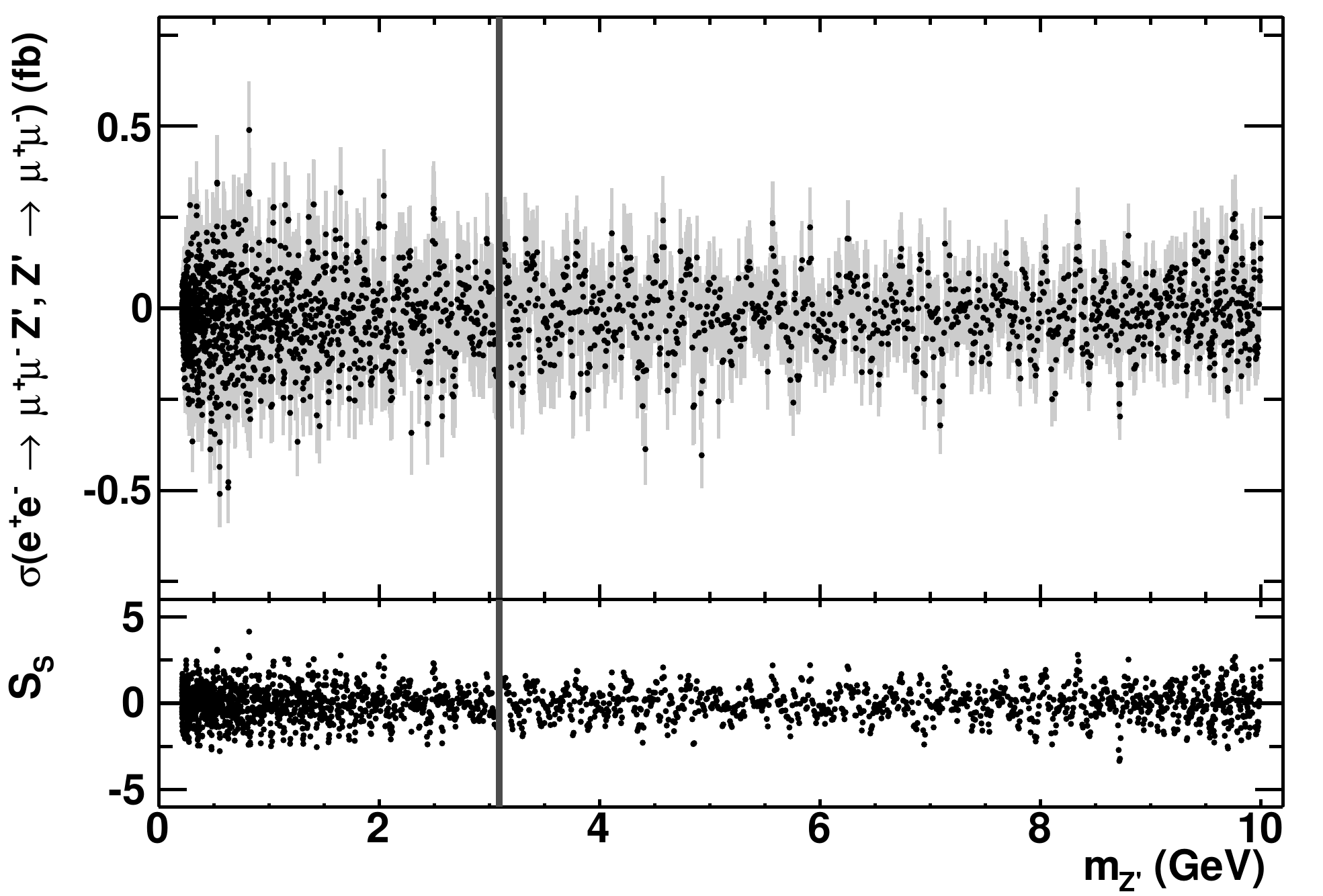}
\caption{The measurement of $e^+ e^- \to \mu^+ \mu^- Z'$, $Z' \to \mu^+ \mu^-$ cross section with its statistical
significance as a function of the $Z'$ mass. The excluded region is indicated by the gray band.}
\label{fig-03}
\end{figure}

\begin{figure}
\centering
\includegraphics[width=8.6cm,clip]{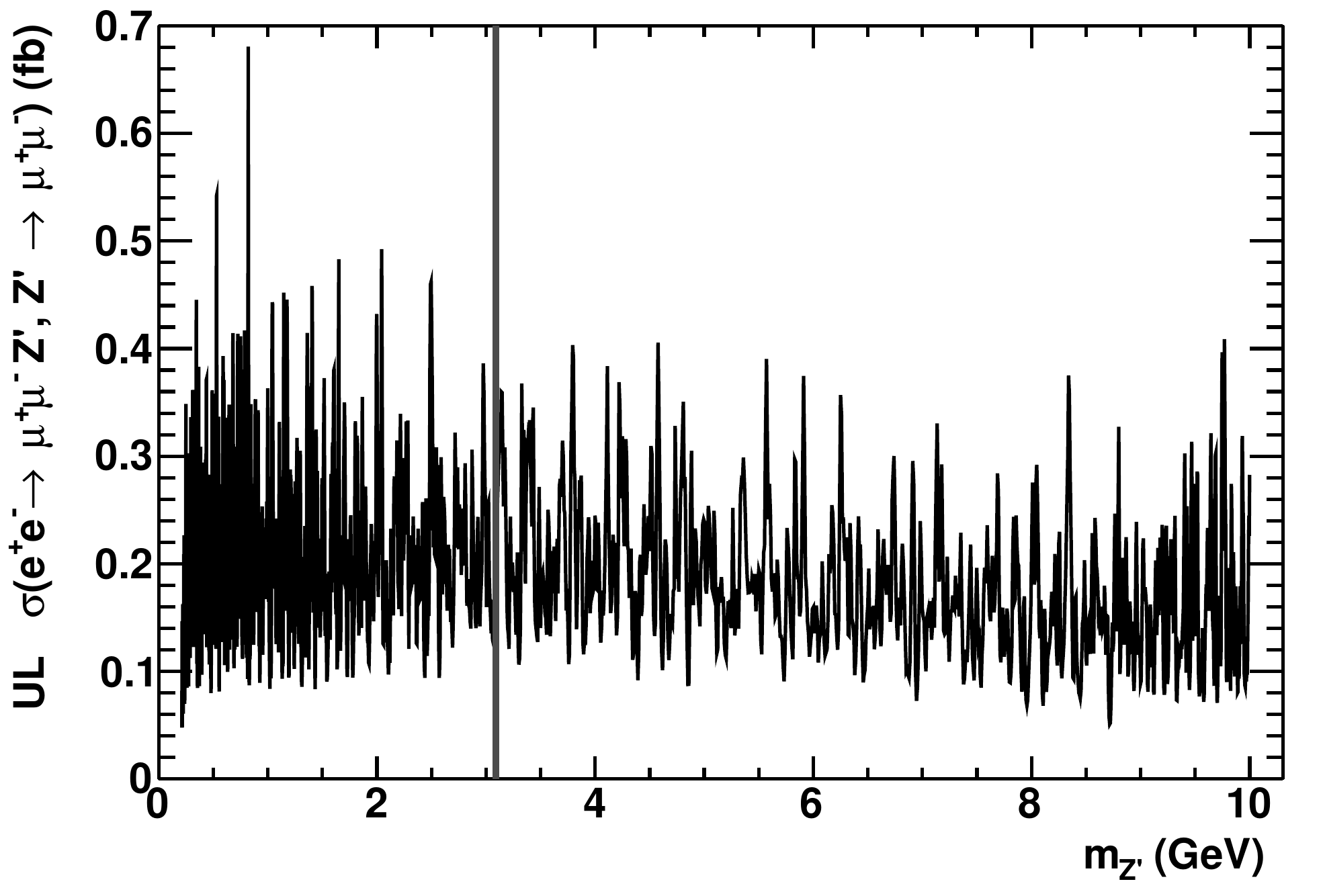}
\caption{The limit on the cross section $\sigma (e^+ e^- \to \mu^+ \mu^- Z'$, $Z' \to \mu^+ \mu^-)$  as a function 
of the $Z'$ mass. The excluded region is indicated by the gray band.}  
\label{fig-04}    
\end{figure}

We also show the distribution of the reduced dimuon mass. The reduced dimuon mass is calculated using the following equation
$m_R = \sqrt{m^2_{\mu^+ \mu^-} - 4m^2_\mu}$ in linear scale as shown in Fig.~\ref{fig-2a} and in log scale as shown
in Fig.~\ref{fig-2b}. The most dominant samples is coming from the direct decay of 
$e^+ e^- \to \mu^+ \mu^- \mu^+ \mu^-$ process. The contribution from the decay of $\Upsilon(2S) \to \pi^+ \pi^- J/\psi$, $J/\psi \to 
\mu^+ \mu^-$ as shown around 3 GeV. The reduced dimoun mass distribution has a better behavior near threshold and it is also
easier to model compare to the dimuon mass distribution.  
The signal efficiency at low masses is about 35\% and it rises to about 50\% around higher mass of the reduced dimuon mass. 
We exclude the $J/\psi$ region when calculating the correction factors by fitting the simulated and reconstructed reduced 
dimuon masses in the range of $1 < m_R < 9$ GeV. We obtain a correction factor of 0.82 as shown in Fig.~\ref{fig-2a}. 
It includes the ISR emission and differences in the trigger efficiency, charged particle identification, and track and photon
reconstruction efficiency. We assign a systematic uncertainty of 5\% to cover the small variations between the uncertainties 
on the $e^+ e^- \to \mu^+ \mu^- \mu^+ \mu^-$ and data taking period. We calculate the ISR contribution based on the quasi real electron
approximation~\cite{benayoun}. We assess the side bands of the four dimuon mass distribution in the range of 5.0 - 8.0 GeV. In this region
the process of $e^+ e^- \to \tau^+ \tau^- (\gamma)$ is dominant. The correction factors are in agreement with the correction factors 
obtain from the reduce dimuon mass spectrum.   
The signal yield is extracted bu a series of unbinned likelihood fits to the reduced dimuon mass spectrum within the range of $0.212 < m_R < 10$ GeV 
and $0.212 < m_R < 9$ GeV for the $\Upsilon(4S)$ resonance data and $\Upsilon(2S)$ and $\Upsilon(3S)$ resonances data, respectively.  
We exclude a region of $\pm 30$ MeV around the nominal known $J/\psi$ mass. We probe a total of 2219 mass hypothesis. The cross section of
$e^+ e^- \to \mu^+ \mu^- Z'$, $Z' \to \mu^+ \mu^-$ is extracted as a function of $Z'$ mass as shown in Fig.~\ref{fig-03}. The gray band indicates 
the excluded region. We find the largest local significance is $4.3 \sigma$ around $Z'$ mass of 0.82 GeV that is corresponding to the global
significance of $1.6 \sigma$ and it is consistent with the zero-hypothesis.  
We also derive $90\%$ confidence level (CL) Bayesian upper limit on the cross section of $e^+ e^- \to \mu^+ \mu^- Z'$, $Z' \to \mu^+ \mu^-$ as shown in
Fig.~\ref{fig-04}. 

\begin{figure}
\centering
\includegraphics[width=9cm,clip]{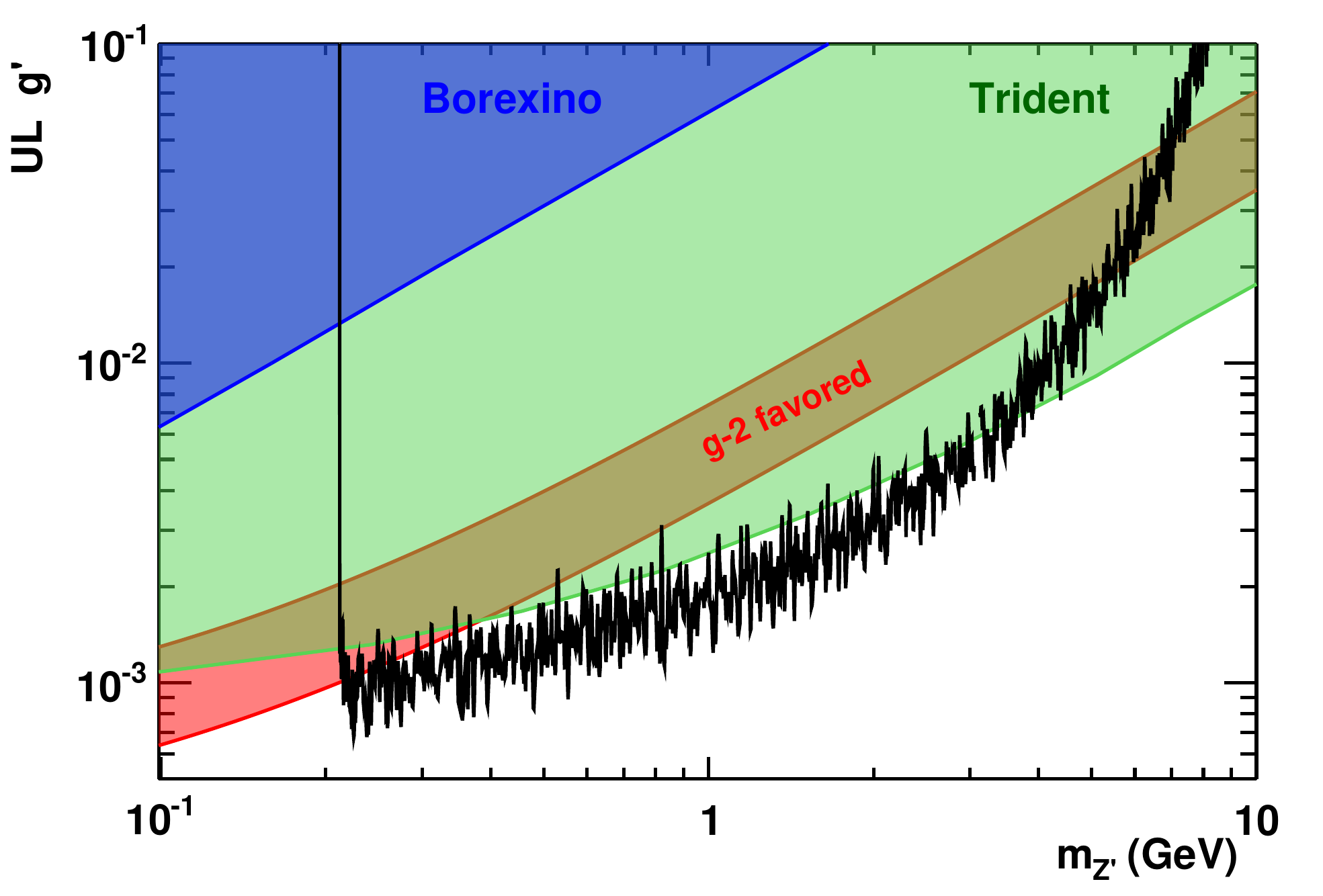}
\caption{Upper limit on the new gauge coupling $g'$ as a function of the mass of $Z'$ together with
the existing limits from Borexino and neutrino experiments.}
\label{fig-05}      
\end{figure}

We consider all uncertainties to be uncorrelated except for the uncertainties of the luminosity and efficiency. We finally extract
the corresponding $90 \%$ CL on the coupling parameter $g'$ by assuming the equal magnitude vector couplings muons, taus and the corresponding
neutrinos together with the existing limits from Borexino and neutrino experiments as shown in Fig.~\ref{fig-05}.  
We set down to $7 \times 10^{-4}$ near the dimuon threshold.

\newpage

\section{Conclusion}
In conclusion, we have performed the first direct measurement of $Z'$ production from the decay of $e^+ e^- \to \mu^+ \mu^- Z'$, 
$Z' \to \mu^+ \mu^-$ an $e^+ e^-$ collider at BABAR.  No significant signal is observed for $Z'$ masses in the range of
0.212 - 10 GeV. We set limits on the coupling parameters $g'$ down to $7 \times 10^{-4}$. We set a strongest bounds for many parameter 
space below 3 GeV. We exclude most of the remaining parameter space preferred by the discrepancy between the calculated and measured 
anomalous magnetic moment of the muon above the dimuon threshold~\cite{echenard}.

\section{ACKNOWLEDGMENTS}
The author would like to thank B. Echenard for the helpful conversations. The author also thanks the organizers of the XLVI 
International Symposium on Multiparticle Dynamics, Jeju Island, South Korea. The support from the BABAR Collaboration, 
the University of South Alabama, and the University of Mississippi is gratefully acknowledged. This work was 
partially supported by the U.S. Department of Energy. 

%

\end{document}